\documentclass[aps,letterpaper,amssymb,showkeys,10pt]{revtex4}
\usepackage{amsmath}

\usepackage{epsfig}

\begin{document}

\title{Cluster Approximation for the Farey Fraction Spin Chain}

\author{Thomas Prellberg}
\email{t.prellberg@qmul.ac.uk} 
\affiliation{School of Mathematical Sciences, 
Queen Mary, University of London,
Mile End Road, London E1 4NS, United Kingdom}
\author{Jan Fiala}
\email{jfiala@clarku.edu}
\affiliation{Department of Physics,
Clark University, Worcester, MA 01610, USA}
\author{Peter Kleban}
\email{kleban@maine.edu}
\affiliation{LASST and Department of Physics \& Astronomy,
University of Maine, Orono, ME 04469, USA}

\date{\today}
\begin{abstract} 
We consider the Farey fraction spin chain in an external field $h$. Utilising ideas from dynamical
systems, the free energy of the model is derived by means of an effective cluster energy
approximation. This approximation is valid for divergent cluster sizes, and hence appropriate for
the discussion of the magnetizing transition. We calculate the phase boundaries and the scaling of
the free energy. At $h=0$ we reproduce the rigorously known asymptotic temperature dependence of the free energy.  For $h \ne 0$, our results are largely consistent with those found previously using mean field 
theory and renormalization group arguments. 
\end{abstract}
\maketitle

\section{Introduction}
\label{section1}

The Farey fraction spin chain, which we study here, is one example of a set of closely related one--dimensional models (see \cite{FK} and  \cite{FK2} for details) which are of interest in both statistical mechanics and number theory.  From the statistical mechanics point of view, there are a number of results, mainly for $h=0$.  All the models have the same free energy $f(\beta,h=0)$, and exhibit a phase transition, at finite temperature, that  is rather unusual. It is known, rigorously, to lie on the border between first- and second-order with the asymptotic form $f(\beta,0) \sim t/\log t$ (where $t=1-\beta/\beta_c$).  In the low--temperature state $f(\beta,0)=0$ and the magnetization is saturated, so there are no thermal effects at all.  For $\beta < \beta_c$, $f(\beta,0)<0$ and the magnetization vanishes (see \cite{FKO} for details).  In \cite{FK} it is proven that the saturated state persists for $h \ne 0$ and $\beta > \beta_c$; when $h >0$, the magnetization $m=1$ and when $h<0$, $m=-1$.  The interesting question is how these states relate to the high--temperature state.  In \cite{FK} this question is addressed via a renormalization group calculation which finds, among other results, a phase diagram that is the same as illustrated in  Figure~\ref{fig3} below.    However, these models have long-range interactions, so the applicability of the renormalization group might be questioned.  It was in fact the desire to verify the results at non-zero $h$ in \cite{FK}  that motivated this work (the results of \cite{FK} for $h=0$ agree with the rigorous behavior).

The Farey fraction spin chain has also led to some new results in number theory   \cite{KOPS, Pe, B}.  Additionally, as  explained below, there is a connection to dynamical systems--in fact, the asymptotic form of $f(\beta,0)$ mentioned comes from a result for dynamical systems \cite{P1}.  Furthermore, chaotic behavior is even exhibited by certain statistical quantities.  In particular, \cite{Pe} proves that the ``density of states" for the infinite chain does not exist--it is a distribution.

In what follows, we analyze the Farey fraction spin chain in a particular approximation, in which
the energy of a configuration is described by single cluster energies. Within this
approximation, the model becomes exactly solvable (and is closely related to the ``necklace" or ``bead" models of Fisher and Felderhof  \cite{F}).  Furthermore, our results for $h \ne 0$ agree, in the main, with a previous analysis \cite{FK} that makes use of mean field theory and the renormalization group.   There are, however, some intriguing differences.

The Farey fraction spin chain \cite{KO,FKO} may be defined as a chain of 
$N$ spins $\sigma_i$; $ i=1,2, \ldots N$ with two possible 
states $\sigma_i\in\{\uparrow,\downarrow\}$. Using the matrices
\begin{equation}\label{dom}
A_{\uparrow}=\begin{pmatrix}1&0\\1&1\end{pmatrix}
\quad\text{and}\quad
A_{\downarrow}=\begin{pmatrix}1&1\\0&1\end{pmatrix}
\;,
\end{equation}
we define the energy of a configuration of $N$ spins $\{\sigma_i\}$ as
\begin{equation}\label{ener}
E_N(\{\sigma_i\},h)=\log \Bigl( {\rm Tr} \prod_{i=1}^NA_{\sigma_i}\Bigr)
-h\sum_{i=1}^N\left(\chi_\uparrow(\sigma_i)-\chi_\downarrow(\sigma_i)\right),
\end{equation} 
where $\chi_\uparrow(\sigma_i)= 1\;(0)$ for $\sigma_i =\;\uparrow (\downarrow)$ so that it counts the number of up spins; $\chi_\downarrow(\sigma_i)$ is defined similarly to count the number of down spins.  The cyclic invariance of the trace in (\ref{ener}) makes the system translationally invariant. The partition function is then given as a sum over $2^N$ spin configuration
\begin{equation}\label{part}
Z_N(\beta,h)=\sum_{\{\sigma_i\}}e^{-\beta E_N(\{\sigma_i\},h)}\ .
\end{equation} 
Our focus is the limiting free energy
\begin{equation}\label{free}
-\beta f(\beta,h)=\lim_{N\rightarrow\infty}\frac1N\log Z_N(\beta,h)\;.
\end{equation} 

The paper is organized as follows.
In section \ref{section2} we utilize the thermodynamic formalism to give a dynamical
systems interpretation of the Farey fraction spin chain and describe how 
this  connection may be used to obtain an effective cluster approximation. 
The analysis of the Farey fraction spin chain within the cluster
approximation is then described in section \ref{section3}, leading to explicit equations for the free energy and
the phase boundaries. 
Section \ref{section4} contains the calculation of the scaling properties near the critical point. A summary and comparison with the results from \cite{FK} are contained in section \ref{section5}. Appendix \ref{appendix1} contains the derivation of the 
asymptotics of the cluster partition function.

The remainder of this section deals with a reformulation of the model in terms of
clusters of consecutive spins of equal state and closes with a brief discussion of our strategy for calculating $f(\beta,h)$ in the cluster approximation.

Iteration of the matrices $A_{\uparrow}$ and $A_{\downarrow}$ leads to
\begin{equation}\label{ite}
A_{\uparrow}^n=\begin{pmatrix}1&0\\n&1\end{pmatrix}
\quad\text{and}\quad
A_{\downarrow}^n=\begin{pmatrix}1&n\\0&1\end{pmatrix}\;.
\end{equation}
One notices that while some matrix elements increase in size, the zero field energy for the associated configurations remains constant, as 
${\rm Tr} A_{\uparrow}^n={\rm Tr} A_{\downarrow}^n=2$. These two states, in fact, are the ground states at zero field.  (They are also responsible for the low temperature thermodynamics \cite{KO}.)
 
The energy is increased considerably,
however, once a {\em change of spin} occurs. It therefore is useful to think of a general configuration as a sequence of clusters of consecutive
 spins of equal state (irrespective of whether the state is $\uparrow$ or $\downarrow$, as in zero field the energy is invariant under spin flip).
 If there is no change of spin at all, one has (as mentioned) 
$Tr(\prod_{i=1}^NA_{\sigma_i})=2$. Once there is a change of spin, we can take advantage of the cylic
invariance of the trace to make the configuration begin with $A_{\uparrow}$ (resp.~$A_{\downarrow}$)
 and end with $A_{\downarrow}$ (resp.~$A_{\uparrow}$). Thus
the total number of spin changes $2K$ must be even (with $K \ge 1$), and we can describe such a spin configuration $\{\sigma_i\}_{i=1}^N$ by a sequence of 
$2K$ clusters of size $n_k\geq1$ with $\sum_{k=1}^{2K}n_k=N$. Therefore, using
\begin{equation}\label{adr}
A_{\downarrow}=SA_{\uparrow}S^{-1}
\quad\text{with}\quad
S=\begin{pmatrix}0&1\\1&0\end{pmatrix}=S^{-1}
\end{equation}
we can write for any configuration with $K\geq1$
\begin{equation}\label{trcl}
{\rm Tr}\prod_{i=1}^NA_{\sigma_i}={\rm Tr}\prod_{k=1}^{2K}M_{n_k}
\quad\text{with}\quad
M_{n_k}=A_{\uparrow}^{n_k}S=\begin{pmatrix}0&1\\1&n_k\end{pmatrix}\;.
\end{equation}

Let us now suppose that we could find a meaningful approximation of the form
\begin{equation}\label{app}
{\rm Tr}\prod_{k=1}^{2K}M_{n_k}\approx\prod_{k=1}^{2K}e^{\epsilon_{n_k}}\;.
\end{equation}
In this case, the energy simplifies considerably and (for all but the ground states) we can write
\begin{equation}\label{appe}
E_N(\{\sigma_i\},h)\approx\sum_{k=1}^{2K}\left(\epsilon_{n_k}-(-1)^{k-1}hn_k\right).
\end{equation}
Working in the grand canonical ensemble, we show in section \ref{section3} that one obtains from this approximation an
exact expression for the limiting free energy $f(\beta,h)$, given by
\begin{equation}\label{fr}
\Lambda(e^{\beta(f-h)},\beta)\Lambda(e^{\beta(f+h)},\beta)=1\;,
\end{equation}
with the cluster generating function
\begin{equation}\label{gf}
\Lambda(z,\beta)=\sum_{n=1}^\infty z^ne^{-\beta\epsilon_n}\;.
\end{equation}
It follows from the approximation discussed in section
\ref{section3} that this generating function has a radius of convergence of $1$. Hence for $\beta > \beta_c(h)$
the limiting free energy is $f(\beta,h)=-|h|$, which for $\beta > \beta_c(0)$ agrees with the rigorously known result in \cite{FK}. Thus
the phase boundary $\beta_c(h)$ is given by 
\begin{equation}\label{pb}
\Lambda(1,\beta_c)\Lambda(e^{-2\beta_c|h|},\beta_c)=1\;.
\end{equation}
In particular, for zero field $h=0$, the free energy $f(\beta,0)$ and critical temperature 
$\beta_c(0)$ follow from
\begin{equation}\label{cc}
\Lambda(e^{\beta f},\beta)=1\quad\text{and}\quad\Lambda(1,\beta_c)=1\;,
\end{equation}
respectively.

\section{Thermodynamic Formalism and Cluster Approximation}\label{section2}

In order to proceed further, we  consider the
thermodynamics of the Farey tree \cite{FPT}. This can be recast in a transfer operator
formulation associated with the iteration of an interval map (see \cite{FK} and \cite{P}). By modifying
this interval map we arrive at the desired cluster approximation.  Note that the Farey tree is known, rigorously, to have the same free energy $f(\beta,h=0)$ as the Farey fraction spin chain (with no external field) \cite{FKO}.

\begin{figure}[ht]
\centering\includegraphics[width=11.0cm]{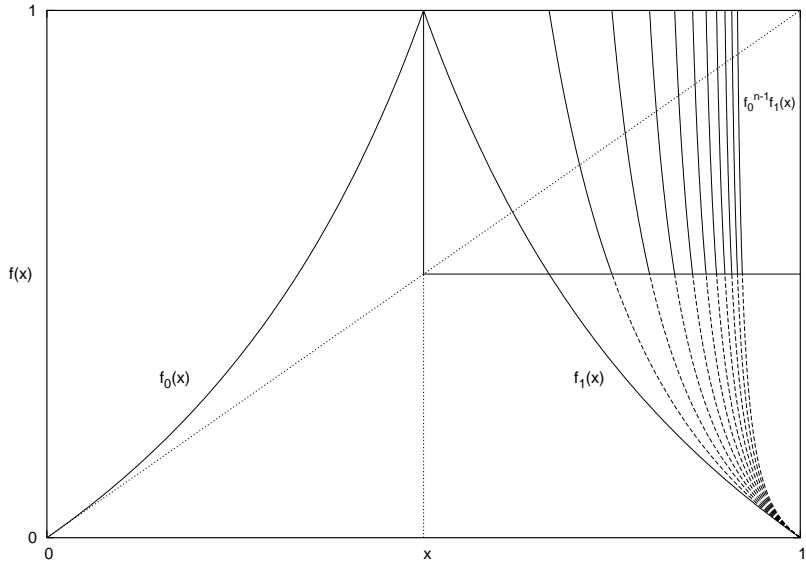}
\caption{Farey map and first-return map on the interval $[1/2,1]$. The first-return map is given
explicitly by the branches ${f_0}^{n-1}f_1$ for $n\in\mathbb N$. Their extension to all of $[1/2,1]$ is shown by dashed lines.}\label{fig1}
\end{figure}

The Farey tree is generated by the Farey map defined on the unit interval $[0,1]$, 
which is defined as
\begin{equation}\label{fm}
f(x)=\left\{\begin{array}{c}
f_0(x)=x/(1-x)\;,\quad\mbox{if $0\leq x\leq1/2$,}\\
f_1(x)=(1-x)/x\;,\quad\mbox{if $1/2<x\leq1$,}
\end{array}\right.
\end{equation}
(see Figure~\ref{fig1}).
We denote the inverses by $F_0(x)={f_0}^{-1}(x)=x/(1+x)$ and $F_1(x)={f_1}^{-1}(x)=1/(1+x)$.
The associated transfer operator is formally given by
\begin{eqnarray}\label{to}
{\cal L}_\beta\;\phi(x)&=&
|{F_0}'(x)|^\beta\phi(F_0(x))+|{F_1}'(x)|^\beta\phi(F_1(x))\nonumber\\
&=&\frac1{(1+x)^{2\beta}}\left[\phi\left(\frac x{1+x}\right)+\phi\left(\frac1{1+x}\right)\right]\;.
\end{eqnarray}
Therefore, the $N$-fold iterated operator ${\cal L}_\beta^N\phi(x)$ consists of $2^N$ terms of the form
\begin{equation}\label{term}
|(F_{\tau_1}\circ F_{\tau_2}\circ\ldots\circ F_{\tau_N})'(x)|^\beta
\phi(F_{\tau_1}\circ F_{\tau_2}\circ\ldots\circ F_{\tau_N}(x))
\end{equation}
with $\tau_i\in\{0,1\}$. 
As we are dealing with iterations of M\"obius transformations of the
form $\frac{ax+b}{cx+d}$ with determinant $\pm1$, we can alternatively 
consider multiplication of the associated matrices. (In a slight abuse of notation,
we shall denote the M\"obius transformation and the associated matrix by the same symbol.) 
We find for instance
\begin{eqnarray}\label{sc}
{\cal L}_\beta^N1(0)=\sum_{\{\tau_i\}}d_{\{\tau_i\}}^{-2\beta}\;,
\end{eqnarray}
where $d_{\{\tau_i\}}$ is just the bottom right entry of the matrix product
\begin{equation}\label{nm}
\prod_{i=1}^NF_{\tau_i}\quad\text{where}\quad
F_0=\begin{pmatrix}1&0\\1&1\end{pmatrix}
\quad\text{and}\quad
F_1=\begin{pmatrix}0&1\\1&1\end{pmatrix}\;.
\end{equation}
Now
\begin{equation}\label{f0f1}
F_0^{n-1}F_1=\begin{pmatrix}0&1\\1&n\end{pmatrix}=M_n
\end{equation}
which immediately suggests a cluster approximation analogous to that for the Farey model discussed in section \ref{section1}. 

There are differences with the Farey
fraction spin chain, however.
One of them  concerns details in the allowed clusterings. 
Both the Farey fraction spin chain partition function and the 
$N$-fold iterated transfer operator are expressible in sums containing $2^N$
terms. However, in the clustering representation of the Farey fraction spin chain 
one needs to take the cyclicity of the trace into account. This means
that in the Farey fraction spin chain all excited states contain an even number of 
clusters. These issues (including multiplicity) are dealt with in section \ref{section3}.

More importantly, the energies are different. Note that the matrices $F_0$ and $F_1$ used to generate the $d_{\{\tau_i\}}$ in (\ref{sc})
 differ from the matrices $A_{\uparrow}$ and $A_{\downarrow}$ employed in the Farey fraction spin chain. Therefore $d_{\{\tau_i\}}$
in (\ref{sc}) cannot be compared directly with the trace
(which is given in a similar way by $a_{\{\sigma_i\}}+d_{\{\sigma_i\}}$) in (\ref{ener}).  Despite this, both expressions lead to the same free energy $f(\beta,h=0)$.
  This was already known via the argument in \cite{FKO} that the largest eigenvalue of ${\cal L}_\beta$ is $e^{2\beta f(2 \beta)}$, and the corresponding eigenvector is positive.  More recently,  the connection has been shown to be more direct.  \cite{FK2} proves that (see (\ref{sc}))  ${\cal L}_\beta^N1(0)=2Z_{N-1}^K(2 \beta)$,
 where $ Z_{N}^K$ is the partition function of the Knauf spin chain, which is rigorously known \cite{FKO} to have the same free energy
 $f(\beta,h=0)$ as the Farey fraction spin chain (with no external field).  Therefore it is reasonable to use the Farey tree cluster energies (see (\ref{ple})) in a cluster approximation for the Farey fraction spin chain.

\begin{figure}[ht]
\centering\includegraphics[width=11.0cm]{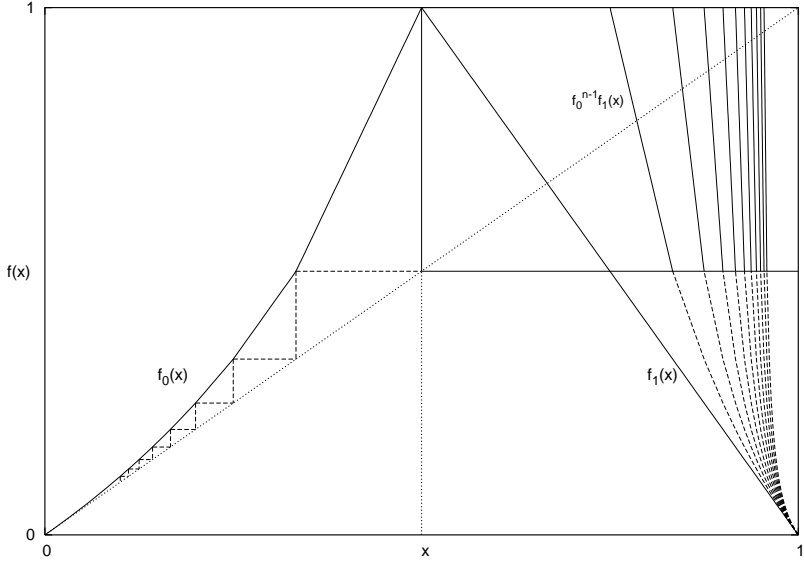}
\caption{Linearized Farey map and first-return map on the interval $[1/2,1]$. The first-return map is given
explicitly by the branches ${f_0}^{n-1}f_1$ for $n\in\mathbb N$, which are linear maps onto $[1/2,1]$. Their extension to all of $[1/2,1]$ is shown
in dashed lines.}\label{fig2}
\end{figure}

We now come to the main point of this section, that one can construct a piecewise linear version
of the Farey map which captures its essential features while being significantly easier to analyse.
This is done by linearizing the map between the inverse images 
$F_0^k(1/2)$ 
(see Figure~\ref{fig2}).
This sequence tends to zero, so that the structure of the map near the
fixed point at zero is preserved under linearization. Due to this fact the critical dynamical
properties of the linearized map and the Farey map are still closely related \cite{GW}. In fact,
the spectral radius of both associated transfer operators shows the same type of
singular behavior at $\beta_c$ \cite{PS}.

Next, we consider the first-return map
on the interval $J=[1/2,1]$, which is the map obtained by repeated iteration of $f$ until $f^n(x)$ lies again in $J$. This map is the key to understanding intermittency in the Farey map \cite{R}. Furthermore, it is responsible for making  the particular piecewise linear map that we use easier to analyse than the original smooth one. 
If we define $n(x)=\min\{n\geq1|f^n(x)\in J\}$ then the first-return map can be written as $g(x)=f^{n(x)}(x)$. Due to our particular choice of $J$
the first-return map $g$ becomes particularly simple. Its branches are given by ${f_0}^{n-1}f_1$ for $n\in\mathbb N$ (see Figure~\ref{fig1}).
One can define a (suitably modified) transfer operator ${\cal M}_{z,\beta}$ for the first-return map, given by
\begin{equation}\label{inducedoperator}
{\cal M}_{z,\beta}\phi(x)=\sum_{n=1}^\infty z^n| (F_1{F_0}^{n-1})' (x)|^\beta\phi(F_1{F_0}^{n-1}(x))\;.
\end{equation}
As the first-return map is only defined on $J$, this operator acts on functions with domain $J$. However, 
using the conjugacy ${\cal C}\phi(x)=|{f_1}'(x)|^\beta\phi(f_1(x))$ we obtain an equivalent conjugate operator 
${\cal C}^{-1}{\cal M}_{z,\beta}{\cal C}$ acting on functions with domain $[0,1]$ and given by
\begin{equation}\label{conjugateinducedoperator}
{\cal C}^{-1}{\cal M}_{z,\beta}{\cal C}\phi(x)=\sum_{n=1}^\infty z^n| ({F_0}^{n-1}F_1)' (x)|^\beta\phi({F_0}^{n-1}F_1(x))\;,
\end{equation}
which allows us to make the identification with $M_n={F_0}^{n-1}F_1$.

The crucial observation is that the operator spectra of ${\cal L}_\beta$ and ${\cal M}_{z,\beta}$ are related \cite{PS,P1}, in the sense that
$\lambda=z^{-1}$ is an eigenvalue of ${\cal L}_\beta$ if and only if $1$ is an eigenvalue of ${\cal M}_{z,\beta}$
(for a rigorous formulation see \cite{PS,R}).

The important consequence of the particular linearization chosen is that the piecewise linearised map replaces the first-return map on $J$ by
a first-return map with branches $f_0^{n-1}f_1(x)$ which are linear and onto, with slopes $n(n+1)$. 
It follows that the eigenfunction associated with the leading eigenvalue $\Lambda(z,\beta)$ of
${\cal M}_{z,\beta}$ becomes constant, and that this eigenvalue is given explicitly by
\begin{eqnarray}\label{plgf}
\Lambda(z,\beta)=\sum_{n=1}^\infty\frac{z^n}{(n(n+1))^\beta}\;.
\end{eqnarray}
Note that $\Lambda(z,\beta)$ extends to an analytic function in the complex $z$-plane cut from $1$ to infinity for 
all $\beta$.

Reformulated in terms of approximations to iterates of the transfer operator ${\cal L}_\beta$ (\ref{sc}),
the linearization implies the approximation
\begin{equation}\label{pl}
{\cal L}_\beta^N1(0)\approx\sum_{\sum_{k=1}^{2K}n_k=N}\frac1{(n_k(n_k+1))^\beta}\;,
\end{equation}
and we recognize that the energy has become a sum of cluster energies
\begin{equation}\label{ple}
\epsilon_n=\log n(n+1)\;.
\end{equation}
In other words, the leading eigenvalue (\ref{plgf}) of the modified transfer operator of the linearized Farey map  is identical with the
cluster generating function obtained from (\ref{gf}) and (\ref{ple}). 

We have mapped the thermodynamics of the Farey tree to a dynamical system given by an interval map, and introduced a linearization of this interval map.   
In this way we obtain a  cluster energy approximation to the original energy expression.  This replaces the original model, which is very difficult to work with, with an approximation that is quite tractable. In the next section, we apply this cluster approximation to the Farey fraction spin chain in a field and determine its consequences.

\section{Thermodynamics in the Cluster Approximation}
\label{section3}

As discussed in the introduction, the Farey partition function can be written in terms of the cluster representation $\{n_k\}_{k=1}^{2K}$. 
There are two ground states with all $N$ spins either up or down. The excited
states consist of configurations with $2K$ alternating clusters of spins with length $n_k \geq 1$. This leads to a degeneracy $n_1$, since (as discussed) we insist that the first and last clusters must have opposite spin directions.

Now the first cluster of size $n_1$ has either all spins up or all spins down.
It is therefore convenient to split the partition function into two terms, according to the state of the first spin. If we denote the partition 
function of configurations with $\sigma_1=\uparrow$ by $Z_N^{\uparrow}(\beta,h)$  and the partition function of configurations with $\sigma_1=\downarrow$ by $Z_N^{\downarrow}(\beta,h)$, then $Z_N^\downarrow(\beta,h)=Z_N^\uparrow(\beta,-h)$ and we obtain
\begin{equation}\label{pfu}
Z_N^\uparrow(\beta,h)=\Bigl(2e^{hN}\Bigr)^{-\beta}+
\sum_{\sum_{k=1}^{2K}n_k=N \atop K\in\{1, \ldots ,\lfloor N/2\rfloor\}}
n_1\Bigl(\prod_{k=1}^{2K}e^{(-1)^{k-1}hn_k}\; {\rm Tr}\prod_{k=1}^{2K}M_{n_k}\Bigr)^{-\beta}\;.
\end{equation}
The partition function can then be obtained as
\begin{equation}\label{tpf}
Z_N(\beta,h)=
Z_N^{\uparrow}(\beta,h)+Z_N^{\downarrow}(\beta,h)=
Z_N^{\uparrow}(\beta,h)+Z_N^{\uparrow}(\beta,-h)\;.
\end{equation}

Using the approximation discussed in Section \ref{section2} leads
to
\begin{equation}\label{appf}
Z_N^\uparrow(\beta,h)\approx\Bigl(2e^{hN}\Bigr)^{-\beta}+
\sum_{\sum_{k=1}^{2K}n_k=N}
n_1\prod_{k=1}^{2K}e^{-\beta(\epsilon_{n_k}+(-1)^{k-1}hn_k)}\;.
\end{equation}
where now
\begin{equation}\label{e}
\epsilon_n=\frac12\log n(n+1)\;,
\end{equation}
with the factor $1/2$ arising from (\ref{sc}). (Note that the energy (\ref{e}) of a cluster is, for large $n$, the same as the energy of the lowest excited state of a chain of length $n$.)  
From this, one can determine the limiting free energy
$f(\beta,h)$. Note also  that this approximation does not change $\beta_c$. This is because the change in spectral radius of the transfer operator occurs at the same temperature after linearization \cite{GW,PS}.

Passing to the grand canonical ensemble, we write
\begin{equation}\label{gce}
G^\uparrow(z,\beta,h)=\sum_{N=1}^\infty z^NZ_N^\uparrow(\beta,h).
\end{equation}
$G^\downarrow(z,\beta,h)$ and $G(z,\beta,h)$ are then defined similarly. One finds
\begin{eqnarray}\label{gceu}
G^\uparrow(z,\beta,h)
&\approx&
2^{-\beta}\frac{ze^{-\beta h}}{1-ze^{-\beta h}}+
\sum_{K=1}^\infty\sum_{n_1,\ldots,n_{2K}=1}^\infty
\!n_1\prod_{k=1}^{2K}(ze^{(-1)^k\beta h})^{n_k}e^{-\beta\epsilon_{n_k}}\nonumber \\
&=&
2^{-\beta}\frac{ze^{-\beta h}}{1-ze^{-\beta h}}+\sum_{K=1}^\infty
\sum_{n_1=1}^\infty n_1(ze^{-\beta h})^{n_1}e^{-\beta\epsilon_{n_1}}
\ldots \nonumber \\
&\ldots&
\sum_{n_{2K}=1}^\infty(ze^{\beta h})^{n_{2K}}e^{-\beta\epsilon_{n_{2K}}}
\;.
\end{eqnarray}
As in the introduction, we define the cluster generating function
\begin{equation}\label{cgf2}
\Lambda(z,\beta)=\sum_{n=1}^\infty z^ne^{-\beta\epsilon_n}\;.
\end{equation}
After some transformations we arrive at (note the close resemblance to the results in section 6 of \cite{F}--one difference being that the parameter $\psi$ is a function of $\beta$ here)
\begin{equation}\label{gcer}
G^\uparrow(z,\beta,h)\approx2^{-\beta}\frac{ze^{-\beta h}}{1-ze^{-\beta h}}+
\frac{ze^{-\beta h}\partial_1\Lambda(ze^{-\beta h},\beta)\Lambda(ze^{\beta h},\beta)}
{1-\Lambda(ze^{-\beta h},\beta)\Lambda(ze^{\beta h},\beta)}\;.
\end{equation}
This is perhaps most easily  seen by expanding equation (\ref{gcer}) backwards. The 
denominator in the second term corresponds to the sum over $n_1$ and $n_2$. The 
partial derivative leads to the multiplying factor $n_1$. Expanding the denominator into
a geometric series produces the product over the sums with summation index $n_i$ with 
$i\geq3$.
We now get
\begin{equation}\label{gcet}
G(z,\beta,h)= G^\uparrow(z,\beta,h)+ G^\downarrow(z,\beta,h),
\end{equation}
where $ G^\downarrow(z,\beta,h)= G^\uparrow(z,\beta,-h)$.
The limiting free energy is then given as
\begin{equation}\label{lfe}
\beta f(\beta,h)=\log z_c(\beta,h)\;, 
\end{equation}
where $z_c(\beta,h)$ is the smallest singularity of $G(z,\beta,h)$ for $z$ on the positive real axis.
This singularity is reached at the smallest positive solution $z_c$ of one of the three equations
\begin{equation}
ze^{-\beta h}=1\;,\quad ze^{\beta h}=1\;,\quad\Lambda(ze^{-\beta h},\beta)\Lambda(ze^{\beta h},\beta)=1\;.
\end{equation}
The first two equations correspond to the two fully magnetized phases, where $z_c=e^{-\beta|h|}$ or, in terms of the free energy,
$f=-|h|$. The third equation corresponds to the high-temperature phase. 
In terms of the free energy,
\begin{equation}\label{ie2}
\Lambda(e^{\beta(f-h)},\beta)\Lambda(e^{\beta(f+h)},\beta)=1\;.
\end{equation}
Since the cluster generating function $\Lambda(z,\beta)$ has radius of
convergence $z=1$, the phase boundary is given by $f=-|h|=-h_c$, and $h_c(\beta)$
is determined by 
\begin{equation}\label{iec}
\Lambda(1,\beta)\Lambda(e^{-2\beta h_c},\beta)=1\;.
\end{equation}
The three phases meet at a critical point given by $h=0$ and $\Lambda(1,\beta_c)=1$, i.e. $\beta_c=2$.

\section{Scaling of the Free Energy}
\label{section4}

In this section we calculate the asymptotic behavior of the solutions $f(\beta,h)$  and  $h_c(\beta)$ of  (\ref{ie2}) and (\ref{iec}), respectively, near the critical point, which is  given by 
$h=0$ and $\beta=\beta_c$. For this we need the asymptotic behavior of the cluster partition function for 
$\beta\rightarrow\beta_c$ and $z\rightarrow1$. Introducing the reduced temperature $t=1-\beta/\beta_c$, we find
\begin{eqnarray}\label{lambdaasy}
\Lambda(z,\beta)\sim1+Ct+(1-z)\log(1-z)
\end{eqnarray}
(a derivation is given in Appendix \ref{appendix1}).
Inserting $z=e^{\beta(f\pm h)}$ into (\ref{lambdaasy}) gives $1-z\sim-\beta_c(f\pm h)$, and we obtain to leading order
\begin{equation}
\Lambda(e^{\beta(f\pm h)},\beta)\sim1+Ct-\beta_c(f\pm h)\log[-\beta_c(f\pm h)]\;.
\end{equation}
Thus, (\ref{ie2}) implies that to leading order
\begin{equation}\label{eqnasy}
2Ct\sim\beta_c(f+h)\log[-\beta_c(f+h)]+\beta_c(f-h)\log[-\beta_c(-f+h)]\;.
\end{equation}
In the field free case ($h=0$) this simplifies to
\begin{equation}
Ct\sim\beta_cf\log(-\beta_cf)\;,
\end{equation}
which can be inverted to give
\begin{equation}\label{zeroh}
f\sim\frac C{\beta_c}\frac t{\log t}\;.
\end{equation}

\begin{figure}[ht]
\centering\includegraphics[width=6.0cm]{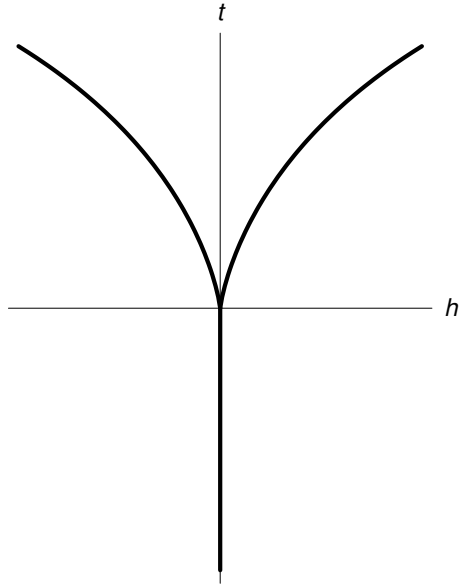}
\caption{Phase boundaries for the Farey fraction spin chain near the critical point, showing the disordered high-temperature phase (top) 
and the two fully magnetized phases (bottom left and right).}\label{fig3}
\end{figure}

For the phase boundary ($-f=|h|=h_c$) we find, using (\ref{iec})
\begin{equation}
2Ct\sim-2\beta_ch_c\log(2\beta_ch_c)\;,
\end{equation}
which can be inverted to give
\begin{equation}\label{hcrit}
h_c\sim-\frac C{\beta_c}\frac t{\log t}\;.
\end{equation}
(see Figure~\ref{fig3}). 
Note that this result implies that on the phase boundary to leading order
$f(\beta,h_c(\beta))\sim f(\beta,h=0)$ as $\beta\rightarrow\beta_c$, and differences only
appear in higher order terms. 

Finally, we compute the leading correction to the zero-field free energy in $h$.
Assuming $|h|\ll-f$ (this assumption of course breaks down near the phase boundary) in (\ref{lambdaasy})
gives
\begin{equation}
2Ct\sim2\beta_cf\log(-\beta_cf)+\beta_ch^2/f\;.
\end{equation}
Inverting this finally leads to
\begin{equation}\label{nonzeroh}
f\sim\frac C{\beta_c}\frac t{\log t}-\frac{\beta_c}{2C}\frac{h^2}t
\end{equation}
for $|h| \ll |t/\log t|$. 

Equations (\ref{zeroh}), (\ref{hcrit}), and (\ref{nonzeroh}) are our main results.  We discuss them in the next section.

\section{Summary and discussion}
\label{section5}

In this paper we have calculated the free energy $f(\beta,h)$ of the Farey fraction spin chain in an external field $h$ by making use of a cluster approximation for the energy of the excited state spin configurations. 

We conclude with a summary of our results and comparison to previous work.
\begin{itemize}
\item In the case of zero field, we find
$$f(\beta,h=0)\sim\frac C{\beta_c}\frac t{\log t}\quad\mbox{where $t=1-\beta/\beta_c$.}$$
The temperature dependence of this result agrees with the known rigorous result \cite{FKO}, the renormalization group calculations \cite{FK}, and a rigorous analysis of the (non-linearized) Farey transfer operator \cite{P}.
\item For the phase boundary, where $h_c=|h|=-f$, we find
$$h_c(\beta) \sim -\frac C{\beta_c}\frac t{\log t}\quad\mbox{where $t=1-\beta/\beta_c$.}$$
The temperature dependence again agrees with the renormalization group calculations \cite{FK}. Additionally, we find that the constants in $f(\beta,0)$ and $h_c(\beta)$ are equal, i.e.~we have to 
leading order $h_c(\beta) \sim -f(\beta,h=0) $, (where $h_c(\beta) = -f(\beta,h_c(\beta))$). This is beyond what renormalization group calculations can predict, since there are several undetermined constants in that case (see \cite{FK} for details). What is more interesting is that this equality is not consistent with the renormalization group results, as explained below.  
\item The change of the free energy for small fields is given by
$$f(\beta,h)\sim\frac C{\beta_c}\frac t{\log t}-\frac{\beta_c}{2C}\frac{h^2}t\quad\mbox{where $t=1-\beta/\beta_c$ and $|h| \ll |t/ \log t|$.}$$
This is not in accordance with renormalization group calculations.  In that case one has \cite{FK}
$$f(\beta,h) \sim a \frac t{\log t}- b \frac{h^2 \log t}t\;,$$
i.e.~a correction term of the order of $h^2\log t/t$, with $a$ and $b$ undetermined constants. Setting this expression equal to $-h_c$ in order to determine the phase boundary, one finds that if the constants in $f(\beta,0)$ and $h_c$ are equal, i.e.~if the cluster results just mentioned hold, one must have $b=0$.  This is, in a sense, consistent, since it might imply that the leading correction to the free energy for finite $h$ is of higher order than $h^2\log t/t$, and therefore could indeed be $h^2/t$.  However, it does not seem to be possible to alter the renormalization calculation to obtain this while keeping the correct form for $f(\beta,0)$, the free energy at $h=0$.  Setting $b=0$ implies that the parameter $x=0$ in \cite{FK}.  If one then includes a higher order term in the flow equation for $u$ (equation (10) in \cite{FK}), the result for $f(\beta,0)$ is no longer correct. 
Another possibility is that $u$ is a ``dangerous irrelevant variable" (or, more
precisely, a ``dangerous marginal variable"), and the finite-field correction
term to the free energy takes the form $h(\ell_0)^2 / (t(\ell_0) \ln
u(\ell_0))$ far from the critical point, which results in a leading
contribution to $f$ in agreement with the cluster result found herein.  Since
very little is understood about $u$, however, such an assumption is completely
{\it ad hoc}.
So the question of the correct leading term for the free energy at finite $h$ remains open. Now, one might question the applicability of the renormalization group to this model, due to the presence of long-range forces. The cluster results, being more closely tailored to the Farey fraction spin chain, are perhaps more compelling, but they are not rigorous either. Therefore, it would be interesting to know what the correct behavior is.  A rigorous asymptotic analysis of the transfer operator for finite $h$ appears possible \cite{P2}, and should answer this question.
\item As mentioned in the Introduction,  there is a set of closely related models, including the  Farey fraction spin chain, which all have the same free energy $f(\beta,h=0)$ at zero external field.  In addition, the magnetization is the same (see \cite{FKO} for details). One therefore expects, according to scaling theory, that they have the same free energy for $h \ne 0$ as well.  The analysis in \cite{FK} makes this assumption.  However, it has not been proven. Given the presence of long-range forces in these systems, one might doubt its validity.  The results obtained here do support it, in that the renormalization group and cluster approximation approaches agree for the most part. However, as mentioned, the agreement is not perfect, and is in any case limited to one particular model (the Farey fraction spin chain).  Therefore further work on these models seems called for.
\end{itemize}

\begin{acknowledgments}
PK acknowledges the hospitality of the Institut f\"{u}r Theoretische Physik, Technische Universit\"{a}t Clausthal, where this work was begun.  We are grateful for discussions with J. Cardy. This research was supported in part by the National Science Foundation Grant No. DMR-0203589.
\end{acknowledgments}

\appendix

\section{Asymptotic Analysis of the Cluster Partition Function}
\label{appendix1}
Here, we present the derivation of the asymptotic behavior (\ref{lambdaasy}) of the cluster partition function
$\Lambda(z,\beta)$ near $(z,\beta)=(1,\beta_c)$, where the critical temperature $\beta_c$ is determined by $\Lambda(1,\beta_c)=1$.

We recall that the cluster partition function is given by
\begin{equation}\label{app1} 
\Lambda(z,\beta)=\sum_{n=1}^\infty z^ne^{-\beta\epsilon_n}\;.
\end{equation}
where in our case
\begin{equation}\label{app2}
\epsilon_n=\frac12\log[n(n+1)]\;,
\end{equation}
and $\Lambda(1,2)=1$ implies $\beta_c=2$.
For $\beta=\beta_c$ one obtains
\begin{equation}
\Lambda(z,\beta_c)=1+\frac{1-z}z\log(1-z)\;.
\end{equation}
For $\beta>1$ and $|z|\leq1$, $\Lambda(z,\beta)$ is an analytic function in $\beta$ with coefficients depending on $z$.
In particular, expanding around $\beta=\beta_c$, we get
\begin{eqnarray}
\Lambda(z,\beta)&=&\Lambda(z,\beta_c)+\sum_{n=1}^\infty\frac{z^n}{n(n+1)}\left([n(n+1)]^{(\beta_c-\beta)/\beta_c}-1\right)\\
&=&\Lambda(z,\beta_c)+\frac{\beta_c-\beta}{\beta_c}\sum_{n=1}^\infty\frac{z^n}{n(n+1)}\log[n(n+1)]+O((\beta_c-\beta)^2)\nonumber
\end{eqnarray}
uniformly in $|z|\leq1$. In particular, letting $z$ approach one from below, the coefficient of the linear term in $(\beta_c-\beta)$ changes continuously with $z$ and we arrive at
\begin{equation}
\Lambda(z,\beta)=\Lambda(z,\beta_c)+\frac{\beta_c-\beta}{\beta_c}(C+o((1-z)^0)+O((\beta_c-\beta)^2)
\end{equation}
with $C=\sum_{n=1}^\infty\frac{\log[n(n+1)]}{n(n+1)}=2.046277452855878591\ldots$ (the sum can easily be evaluated numerically using the Euler-MacLaurin formula).
Introducing the reduced temperature $t=1-\beta/\beta_c$, this implies
\begin{eqnarray}\label{appasy}
\Lambda(z,\beta)\sim1+Ct+(1-z)\log(1-z)\;,
\end{eqnarray}
which is our desired equation (\ref{lambdaasy}).

We conclude this appendix with a few generalizing remarks. For our purposes it was sufficient to work directly with (\ref{plgf}), but we would like to point out that it is possible to perform a more thorough analysis. Observing that
$$
\frac1{[n(n+1)]^{\beta/2}}=\int_0^\infty K(\beta,s)e^{-ns}ds\;,
$$
where
$$
K(\beta,s)=\frac{\sqrt{\pi}}{\Gamma(\beta/2)}I_{(\beta-1)/2}(s/2)e^{-s/2}s^{(\beta-1)/2}\;,
$$
leads to
$$
\Lambda(z,\beta)=\sum_{n=1}^\infty\frac{z^n}{[n(n+1)]^{\beta/2}}=\int_0^\infty K(\beta,s)\frac z{e^s-z}ds\;,
$$
an integral representation which is a different starting point for an asymptotic analysis. 
In particular, one recognizes directly that the large-$n$ asymptotics of the cluster energies $\epsilon_n$ 
is related to the small-$s$ expansion of the integral kernel $K(\beta,s)$, which in turn determines the singular
behavior of $\Lambda(z,\beta)$. The argument can therefore also be extended to more general $\epsilon_n$.


\begin{thebibliography}{99}

\bibitem{FK}
J. Fiala and P. Kleban, {\it Thermodynamics of the Farey fraction spin chain},
J. Stat. Phys. {\bf 110}, 73-86 (2004) [arXiv: math-ph/0310016].

\bibitem{FK2}
J. Fiala and P. Kleban, {\it Generalized number theoretic spin chain--connections to dynamical systems and expectation values}, J. Stat. Phys. to appear [arXiv: math-ph/0503030].

\bibitem{FKO}
J. Fiala, P. Kleban and A. \"{O}zl\"{u}k {\it The phase transition in statistical models defined on Farey fractions}, J. Stat. Phys. {\bf 110}, 73-86 (2003) [arXiv: math-ph/0203048].

\bibitem{KOPS} 
J. Kallies, A. \"Ozl\"uk, M. Peter and C. Snyder, {\it On asymptotic properties of a number
 theoretic function arising from a problem in statistical mechanics} Commun. Math. Phys. {\bf 222}, 9-43 (2001). 

\bibitem{Pe}
M. Peter, {\it The limit distribution of a number-theoretic function arising from
 a problem in statistical mechanics}, J. Number Theory {\bf 90}, 265-280 (2001).
 
\bibitem{B}
F. Boca, {\it Products of matrices   and   and the distribution of reduced quadratic irrationals}, preprint [arXiv: math.NT/0503186].

\bibitem{P1} 
T. Prellberg, {\it Maps of intervals with indifferent fixed points: 
thermodynamic formalism and phase transition}, Ph.D. thesis, Virginia Tech (1991).

\bibitem{F}
M. E. Fisher {\it Walks, walls wetting and melting},
J. Stat. Phys. {\bf 34}, 667-729 (1984).

\bibitem {KO}
P. Kleban and A. \"{O}zl\"{u}k, {\it A Farey fraction spin chain}, Commun. Math. Phys. {\bf 203}, 635-647 (1999).

\bibitem{FPT} 
M. J. Feigenbaum, I. Procaccia, and T. T\'{e}l, 
{\it Scaling properties of multifractals as an eigenvalue problem}, Phys. Rev. A {\bf 39}, 5359-5372  (1989).

\bibitem{P} 
T. Prellberg, {\it Towards a complete determination of the spectrum of a transfer operator associated with intermittency}, J. Phys. A: Math. Gen. {\bf 36}, 2455-2461 (2003).

\bibitem{CK} 
P. Contucci and A. Knauf, {\it The phase transition of the number-theoretic
 spin chain}, Forum Mathematicum {\bf 9}, 547-567 (1997).

\bibitem{GW} 
P. Gaspard and X.-J. Wang, {\it Sporadicity: Between Periodic and Chaotic Dynamical Behaviors}, Proc. Nat. Acad. Sci. USA {\bf 85}, 4591-4595 (1988).

\bibitem{PS} 
T. Prellberg and J. Slawny, {\it Maps of intervals with indifferent fixed points: 
thermodynamic formalism and phase transition}, J. Stat. Phys. {\bf 66}, 503-514 (1992).

\bibitem{R}
H. H. Rugh, {\it Intermittency and regularized Fredholm determinants}, Invent. Math. {\bf 135}, 1-24 (1999).

\bibitem{P2} 
T. Prellberg, {\it private communication}.

\end{thebibliography}
\end{document}